\documentclass[aps,longbibliography,twocolumn,showpacs,superscriptaddress,floatfix,prx]{revtex4-2}
\usepackage{amsfonts,hyperref}

\usepackage{graphicx}
\usepackage{amsmath}
\usepackage{amssymb}

\newcommand{\ZZ}{$\mathbb{Z}_2$ }

\begin{document}
\title{Exploiting the Hermitian symmetry in tensor network algorithms}

\author{Oscar \surname{van Alphen}}  \affiliation{Institute for Theoretical Physics, University of Amsterdam, Science Park 904, 1098 XH Amsterdam, The Netherlands}
\author{Stijn V. \surname{Kleijweg}}\affiliation{Institute for Theoretical Physics, University of Amsterdam, Science Park 904, 1098 XH Amsterdam, The Netherlands}
\author{Juraj \surname{Hasik}}  \affiliation{Institute for Theoretical Physics, University of Amsterdam, Science Park 904, 1098 XH Amsterdam, The Netherlands}
 \affiliation{Department of Physics, University of Zurich, Winterthurerstrasse 190, 8057 Zurich, Switzerland}
\author{Philippe \surname{Corboz}} \affiliation{Institute for Theoretical Physics, University of Amsterdam, Science Park 904, 1098 XH Amsterdam, The Netherlands}

\begin{abstract}
Exploiting symmetries in tensor network algorithms plays a key role for reducing the computational and memory costs.
Here we explain how to incorporate the Hermitian symmetry in double-layer tensor networks, which naturally arise in methods based on projected entangled-pair states (PEPS). For real-valued tensors the Hermitian symmetry defines a \ZZ symmetry on the combined bra and ket  auxiliary level of the tensors. By implementing this symmetry, a speedup of the computation time by up to a factor $4$ can be achieved, while expectation values of observables and reduced density matrices remain Hermitian by construction. Benchmark results based on the corner transfer matrix renormalization group (CTMRG) and higher-order tensor renormalization group (HOTRG) are presented. We also discuss how to implement the Hermitian symmetry in the complex case, where a similar speedup can be achieved.
\end{abstract}

\maketitle

\section{Introduction}
Tensor networks  offer highly accurate numerical tools for the study of strongly-correlated systems, in particular in cases where Monte Carlo methods are ineffective due to the negative sign problem. Most notably, matrix product states have revolutionized the study of quasi-one dimensional systems~\cite{white1992,schollwoeck2011}. Their higher dimensional generalization, known as projected entangled-pair states (PEPS)~\cite{verstraete2004,nishio2004,jordan2008}, have become a state-of-the-art tool for two dimensional systems, with applications ranging from frustrated magnets to strongly correlated electron systems, see e.g. Refs.~\cite{corboz14_shastry, nataf16, liao17, niesen17, chen18, lee18, jahromi18, niesen18, yamaguchi18,kshetrimayum19b, chung19, ponsioen19, lee20, gauthe20, jimenez21, czarnik21, hasik21, shi22, liu22b, gauthe22, peschke22,hasik22, sinha22, ponsioen23b, weerda23, xu23b, hasik24,schmoll24}. The main idea is to efficiently represent quantum many-body states by a contraction of tensors, where the accuracy is systematically controlled by the bond dimension $D$ of the tensors.

A central ingredient to speeding up tensor network calculations is exploiting global symmetries of the states, such as the  U(1) symmetry for systems that conserve the total number of particles~\cite{singh2011,bauer2011}, or non-abelian SU(2) symmetry in certain spin models~\cite{mcculloch02,singh12,bruognolo21,schmoll20}. In the presence of a global symmetry, the tensors can be written in block-sparse form, where the blocks are labeled by the corresponding quantum numbers (e.g. different particle number sectors), and tensor contractions decompose into a product of several smaller blocks, similar to the multiplication of block diagonal matrices, enabling a substantial reduction in computational cost. Another important direction is classifying exotic phases of matter according to the symmetry properties of the tensors in their auxiliary space~\cite{schuch10,Buerschaper2014,duivenvoorden17,bultinck2017,crone20,sahinoglu21}. 

In this paper we consider a different symmetry inherent to standard tensor network calculations, namely the Hermitian symmetry of double-layer networks representing, e.g., the norm of a state or a reduced density matrix (RDM) in the physical (or auxiliary) space. An RDM computed from an MPS is always  Hermitian up to machine precision. However, for PEPS, since the contraction is typically done only approximately, the contraction error can lead to violations of the Hermiticity of RDMs. 

Here we show how to explicitly preserve the Hermiticity by working in a symmetric basis of the combined bra and ket auxiliary space. In the case of real-valued tensors, the Hermitian symmetry corresponds to a $\mathbb{Z}_2$ symmetry which can be implemented in the standard way~\cite{singh2011,bauer2011}, or by making use of symmetric tensor libraries~\cite{fishman2022itensor, hauschild2018efficient, zhai2023block2, tensornetwork2019, wu2023cytnx, tenes, TensorKitJL, weichselbaum24, pepstorch2024, ad-peps, naumann2023varipeps, PEPSKit, TenNetLib, yastn2024}. 

The main benefit, besides preserving Hermiticity, is a roughly fourfold speedup in computation time and a twofold reduction in memory.
As a proof of concept, we present benchmark results based on two standard contraction methods: the corner transfer-matrix renormalization group (CTMRG)~\cite{nishino1996} and the higher-order tensor renormalization group (HOTRG)~\cite{xie12}. We also discuss the complex-valued case, which does not lead to a regular block-sparse form of a tensor, but results in a block-sparse structure for the real and complex parts individually. By exploiting this structure, a similar speedup can be achieved.

\section{Method}

\subsection{iPEPS ansatz and contraction}
\begin{figure}[tb]
  \centering
  \includegraphics[width=\linewidth]{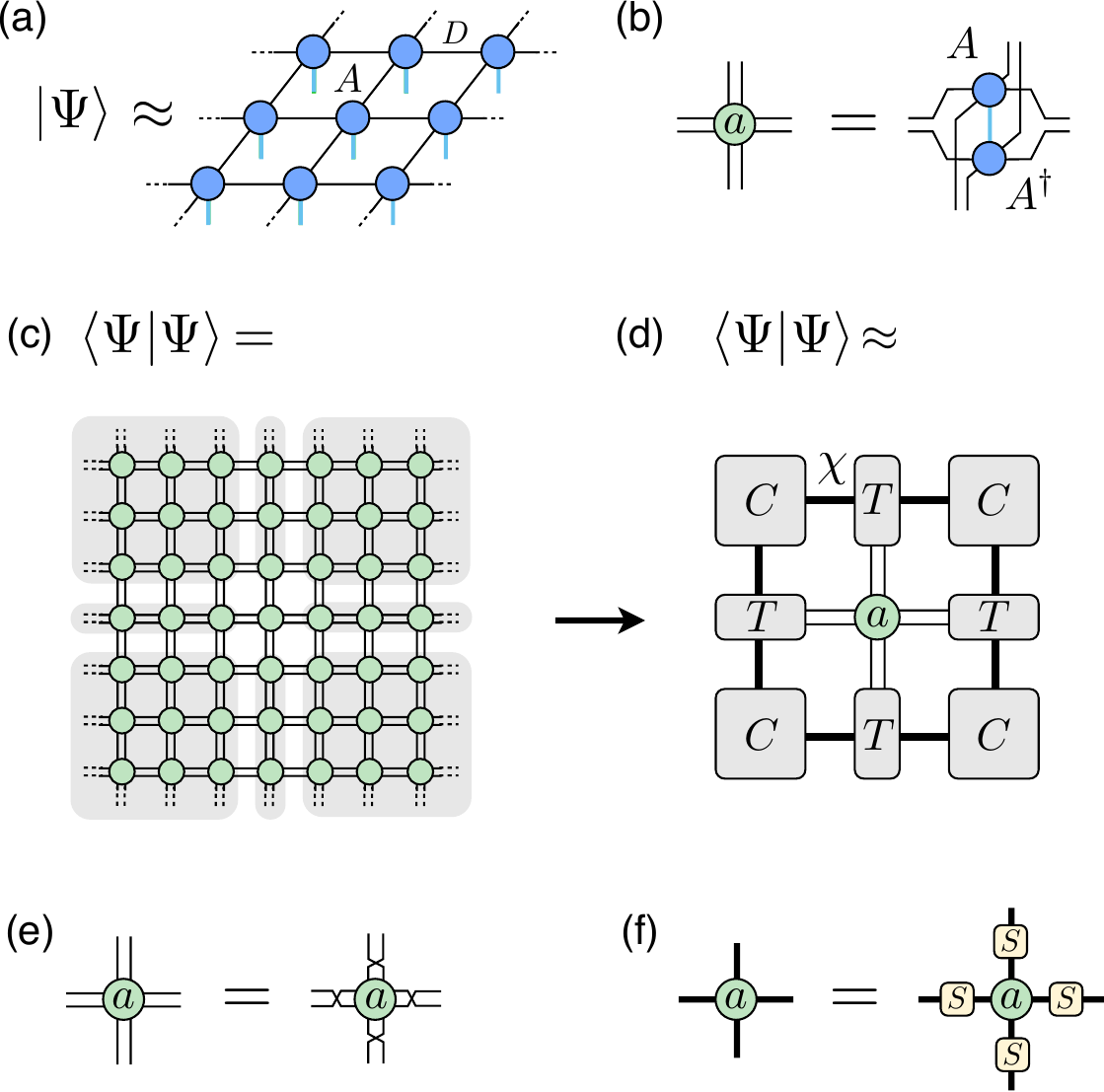}
  \caption{(a) Translationally invariant iPEPS parametrized by a tensor $A$ with bond dimension $D$. The vertical blue legs represent the physical indices on each lattice site. (b)~The double-layer tensor $a$ is defined as the contraction of $A$ with $A^\dagger$ along the physical leg. (c) Representation of the norm in terms of the $a$ tensors. (d) In CTMRG the tensor network surrounding the center tensor $a$ is approximated by four corner ($C$) and edge ($T$) environment tensors, where the accuracy of the approximate contraction is controlled by the bond dimension $\chi$. (e) For real-valued tensors, $a$ is invariant under the exchange of all ket legs with bra legs. (f) The bra and ket  indices can be combined into a bigger index of bond dimension $D^2$. $S$ represents the swap operation of bra and ket  legs shown in~(e). }
  \label{fig:drawings}
\end{figure}

We consider a translationally invariant infinite PEPS (iPEPS)~\cite{jordan2008} which represents a 2D wave function in the thermodynamic limit. It consists of a rank-5 tensor $A$ repeated on a square lattice as shown in Fig.~\ref{fig:drawings}(a). Each tensor has 4 auxiliary legs with bond dimension $D$ connecting  the neighboring tensors and one physical leg with the dimension $d$ of the local Hilbert space of a lattice site. Combining each tensor $A$ with its conjugate $A^\dagger$ into a new tensor $a$ (Fig.~\ref{fig:drawings}(b)) results in a square lattice tensor network shown in Fig.~\ref{fig:drawings}(c), representing the norm of the wave function.

To approximately contract the network, we use CTMRG~\cite{nishino1996} and HOTRG~\cite{xie12}, where the accuracy is controlled by the bond dimension $\chi$. In CTMRG, the system with open boundary conditions is grown in all four directions by iteratively absorbing rows and columns of the network into the boundary tensors with bond dimension $\chi$. Each iteration increases the boundary bond dimension by a factor $D^2$, which is truncated down to $\chi$ based on an eigenvalue or singular value decomposition (SVD)~\cite{nishino1996}. The boundary tensors consist of four corner tensors $C$ and four edge tensors $T$, representing a quarter infinite system or an infinite half-row or half-column of tensors, see Fig.~\ref{fig:drawings}(d). For simplicity, we consider translationally invariant states with a $C4v$ symmetry such that all corners and edges are equivalent. HOTRG is based on iteratively coarse-graining the lattice of tensors. At each iteration, two neighboring $a$ tensors are combined into one tensor (alternating between bra and ket horizontal and vertical direction), followed by a truncation down to the $\chi$ most relevant states based on an eigenvalue decomposition (or SVD)~\cite{xie12}.  

\subsection{\ZZ  symmetry in the real-valued case}
Let us consider the combined bra and ket tensor PEPS in Fig.~\ref{fig:drawings}(b). If we take the tensor $A$ to be real valued, the bra and ket tensors are  equal, i.e. $A = A^\dagger$, and the double-layer tensor $a_{i\bar{i} j\bar{j} k\bar{k} l\bar{l}}$ is invariant under the simultaneous exchange of all bra and ket  indices, i.e.  $a_{i\bar{i} j\bar{j} k\bar{k} l\bar{l}}= a_{\bar{i}i \bar{j}j \bar{k}k \bar{l}l}$, as depicted in Fig.~\ref{fig:drawings}(e). The swapping of the legs can also be represented as a swap tensor $S$, defined as $S_{i \bar{i} j\bar{j}} = \delta_{i \bar{j}} \delta_{\bar{i} j}$.

By combining each pair of virtual indices on the bra and ket  level into one big index of bond dimension $D^2$, we can represent tensor $a$ as a four-legged tensor shown in Fig.~\ref{fig:drawings}(e). In this combined basis $S$ becomes a $D^2 \times D^2$ matrix acting on a joint leg, and acting with $S$ on all four legs of $a$ simultaneously leaves the tensor $a$ invariant (Fig.~\ref{fig:drawings}(f)). Since $S^2 = \mathbb{I}$, $S$ defines a $\mathbb{Z}_2$ symmetry for tensor $a$.

As with other symmetries, if we represent the tensor $a$ in the eigenbasis of $S$, the tensor acquires a block-sparse structure, where the blocks can be labelled by the eigenvalues of $S$, which are either $+1$ (even) or $-1$ (odd). For example for $D=2$, starting from the combined  basis of the bra and ket virtual index $\left\{ (1 \bar{1}),  (2 \bar{1}), (1 \bar{2}), (2 \bar{2})\right\}$, the eigenbasis of $S$ is given by $\left\{ (1 \bar{1}),   (2 \bar{2}), (1 \bar{2} + 2 \bar{1})/\sqrt{2}, (1 \bar{2} - 2 \bar{1})/\sqrt{2}  \right\}$, where the first three states belong to the even sector (eigenvalue +1) and the last state to the odd sector (eigenvalue -1). 
One can easily show that for a given $D$ the even sector has dimension $dim(even) = D + (D^2 -D)/2$ and $dim(odd) = (D^2 -D)/2$. This is because there are $D$ even basis states of the form $(k \bar{k})$, and half of the remaining states is a symmetric combination $(k \bar{l} + l \bar{k})/\sqrt{2}$, the other half an antisymmetric combination $(k \bar{l} - l \bar{k})/\sqrt{2}$, with $k \neq l$, respectively.

Working in the eigenbasis of $S$ has three advantages. (1)~Thanks to the block-sparse structure of the tensors, the computational cost to contract two tensors or to perform an SVD of a tensor is lower than in the dense case. (2)~The block-sparse structure can also be exploited to reduce the memory cost, by storing only the blocks which are non-zero. (3)~The tensor $a$ retains its symmetry also when the virtual space is truncated to a lower bond dimension. The latter is not the case in the non-symmetric basis. For example, for $D=2$ when keeping only the first two  basis states $\left\{ (1 \bar{1}),  (2 \bar{1})\right\}$, the tensor $a$ would no longer be Hermitian. Using  tensors that break the physical Hermitian symmetry will result in non-physical quantities, e.g. any reduced density matrix computed from such tensors will not be Hermitian.

Having identified the underlying $\mathbb{Z}_2$ symmetry of tensor $a$, one can implement the symmetry as described in Refs.~\cite{singh2011,bauer2011} (or use a symmetric tensor  library) to contract tensors and to perform operations on tensors like an SVD or an eigenvalue decomposition. What is different from the standard case of a physical global  $\mathbb{Z}_2$ symmetry (like e.g. in the transverse field Ising model or parity conservation in fermionic systems), is that the even and odd sectors can only be defined on the combined virtual bra and ket  spaces (e.g. for $a$), but not on the individual bra and ket  level (e.g. for $A$).   
As a consequence, for a multiplication of e.g. an $a$ tensor with an $A$ tensor, a dense contraction has to be performed, because the $A$ tensor is not symmetric. Thus, for certain operations involving individual bra- or ket- tensors, the $\mathbb{Z}_2$ symmetry cannot be exploited. Still, typically the computationally most expensive calculations in 2D contraction algorithms are those involving only combined bra and ket  spaces, e.g.  the SVD in CTMRG's renormalization step or in essentially all steps in HOTRG (except in the initialization step in which the $a$ tensor is formed, which is computationally subleading).

The maximal gain in computational cost is a factor 4, which is obtained if the sizes of the even and odd sectors are equal. This can be easily seen from the multiplication of two block-diagonal matrices of size $M\times M$. The computational cost of the dense multiplication scales as $M^3$. Multiplying two blocks of half of the size has a reduced cost of  $2\times(M/2)^3 =  M^3/4$, hence a factor~4. 
For the $a$ tensor $dim(even)=dim(odd)$  is asymptotically true in the infinite $D$ limit. 
For the renormalized spaces in CTMRG and HOTRG, we will analyze the relative sizes of even and odd sectors in the results in Sec.\ref{sec:results}. 
In practice, a factor 4 may not be achieved, even for $dim(even)=dim(odd)$, because keeping track of the sectors comes with a certain computational overhead (depending on the actual implementation), such that a factor 4 is only reached  asymptotically in the large bond dimension limit.

We note that the Hermitian $\mathbb{Z}_2$ symmetry cannot be trivially combined with a global symmetry such as U(1) or SU(2). For example, consider the U(1) symmetry arising from conservation of the total $S^z$ in a spin model. Here, the swap tensor $S$ mixes different sectors of the total $S^z$~\cite{su2comment}. One could in principle still implement the $\mathbb{Z}_2$ for the total $S^z=0$ block, and exploit the fact that the $\mathbb{Z}_2$ Hermitian symmetry results in a block sparse structure, where the blocks in the  total $S^z=k$ can be mapped onto blocks with total $S^z=-k$, so that it is sufficient to keep only the positive sectors. 
 Thus, while profiting from the  $\mathbb{Z}_2$ Hermitian symmetry is possible in this context, it would involve modifications in the underlying tensor library to deal with this particular case, which is beyond the scope of this work.

% It is worth noting that in the $SU(2)$ case for certain choices of virtual spaces the Hermitian $\mathbb{Z}_2$ symmetry is directly included. For example, if one considers an iPEPS with $D=2$ with virtual indices carrying a single spin-1/2 representation, the fusion of bra and ket virtual indices yields $1/2 \otimes 1/2 = 1 \oplus 0$, corresponding to the three even (triplet) and one odd (singlet) states. But more generally, as in the U(1) case, the blocks are not automatically Hermitian symmetric. Consider, e.g. a virtual space with $D=6$ with representations $1/2 \oplus 3/2$. The basis states of $1/2 \otimes 3/2 = 1 \oplus 2$ and $3/2 \otimes 1/2 = 1 \oplus 2$ are not Hermitian symmetric, and additional effort would be required to preserve the symmetry. 

% eigenvectors of the Hermitian Z_2 symmetry operator S

\subsection{Hermitian symmetry in the complex case}

For complex valued tensors, $A \neq A^\dagger$ and hence the $a$ tensor is not invariant upon exchanging all bra and ket  legs. The real part of $a$ is invariant, but the imaginary part of $a$ requires a minus sign due to the complex conjugation. One can nevertheless implement the Hermitian symmetry and obtain up to factor 4 speed up as we explain in the following.

We first recall that multiplying two complex numbers comes with a factor 4 compared to real numbers, because it involves 4 multiplications instead of only a single one, e.g. $(a + b i) (c + d i) = ac - bd + (bc + ad) i$. 
Now, if we represent $a$ in the eigenbasis of $S$ it turns out that all the blocks in the total even sector are purely real valued, whereas all blocks in the total odd sector are purely imaginary. Thus, the $a$ tensor is not block sparse, but it nevertheless has a block structure, where each block contains either only real numbers or imaginary numbers. 
As a consequence, when multiplying two matrices (or tensors) in the eigenbasis of $S$, all operations involve multiplications between purely real numbers and/or purely imaginary numbers requiring only a single multiplication per pair of numbers, hence there is a factor 4 reduction in computational cost. 
In addition, as in the real-valued case, working in the eigenbasis of $S$ has the advantage that truncations on the virtual space will preserve the Hermitian symmetry of the tensor network.

Working with dense tensors with a block structure is typically not supported in standard symmetric libraries. However, one can still make use of standard libraries by using the following trick. Imagine we want to multiply two matrices (or tensors) $b$ which exhibit such a block structure, but which are not block sparse. We can now add an extra leg of dimension 2 with index $q$ to each tensor, resulting in a new tensor $\tilde b$, with $q=1$ and $q=2$ corresponding to the even and odd sector. We can then store the real part and imaginary part of $b$ in the $q=1$ and $q=2$ slices, respectively,
\begin{eqnarray}
\tilde{b}(:,:,1) &=& \mathrm{Re}[b(:,:)],\\ \nonumber
\tilde{b}(:,:,2) &=& \mathrm{Im}[b(:,:)].
\end{eqnarray}
The new tensor $\tilde{b}$ has purely real numbers and a block-sparse structure, i.e. it can be represented using a standard symmetric tensor library. The contraction of the two tensors involves an extra tensor $f$ which combines the real and imaginary parts according to the multiplication rules of complex numbers, i.e.
\begin{eqnarray}
f(1,1,1) &=& 1, \quad f(2,2,1)= -1 \\  \nonumber
f(1,2,2) &=& 1, \quad f(2,1,2)= 1,  
\end{eqnarray}
and the multiplication of two $\tilde{b}$ tensors then reads 
\begin{equation}
\tilde{c}_{ikr} = \sum_{jpq} \tilde{b}_{ijp}\tilde{b}_{jkq} f_{pqr}.
\end{equation}
Converting such a real-valued tensor back to a complex-valued one can simply be done by multiplying the extra leg by the vector $[1, i]$.

\section{Benchmark results}
In this section we present benchmark results for the contraction of the ground state of the Heisenberg model based on the converged tensors from Ref.~\cite{hasik21}, using CTMRG and HOTRG with and without implemented $\mathbb{Z}_2$ symmetry. The lowest energy states for different bond dimensions $D$ allow for a real-valued tensor network representation.

\subsection{Hermitian symmetry error}
As discussed in the previous section, in standard CTMRG and HOTRG the truncation performed during the contraction may be such that the Hermitian symmetry of the tensor network is lost, which may lead to unphysical results, such as non-Hermitian reduced density matrices. For CTMRG we define the following Hermitian symmetry error:
\begin{equation}
S_{err}  = \frac{|| \epsilon - \epsilon^\dagger ||}{|| \epsilon+\epsilon^\dagger ||},
\end{equation}
where $\epsilon$ corresponds to Fig.~\ref{fig:drawings}(d) with the tensor $a$ removed in the center, i.e. a one-site environment with four open auxiliary bonds on the bra and on the ket level, respectively. If the Hermitian symmetry is preserved during the CTMRG iterations, we should obtain $\epsilon=\epsilon^\dagger$ and hence $S_{err}=0$. 
To measure the symmetry error in HOTRG, we replace one of the $a$ tensors in the network by another tensor $b$ which, when reshaped to a $D^4 \times D^4$ matrix, has $1$'s in the upper triangle and $-1$'s in the lower triangle (and zeros on the diagonal). If the environment is Hermitian, the contraction yields zero, and hence deviations from zero imply a non-Hermitian environment. We normalize the error by a similar evaluation of the network using $|b|$, i.e. with all off-diagonal elements being 1 when reshaped to a matrix.

\begin{figure}[tb]
  \centering
  \includegraphics[width=\linewidth]{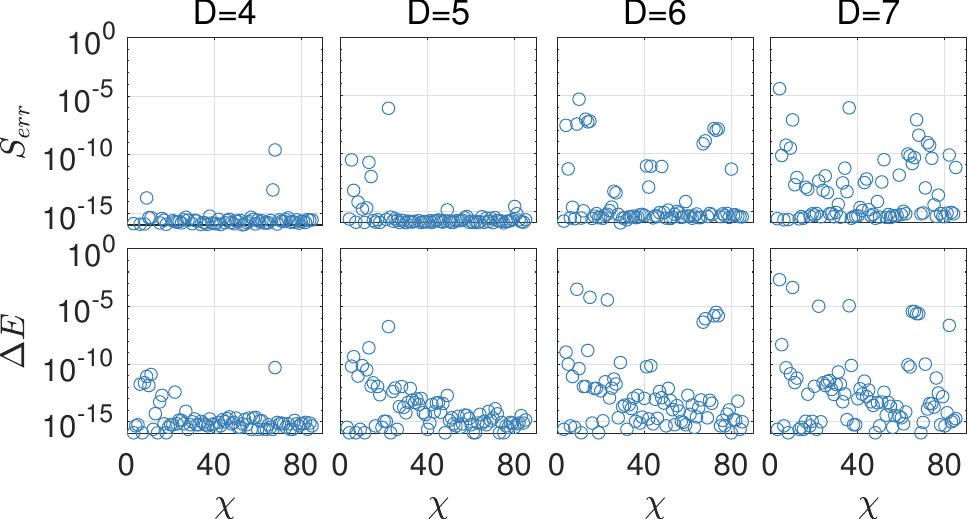}
  \caption{Upper panels: Symmetry error in CTMRG as a function of $\chi$ for different bond dimensions  $D$. Lower panels: Energy difference between the non-symmetric and $\mathbb{Z}_2$ symmetric implementation of CTMRG. }
  \label{fig:ctmrgerror}
\end{figure}

Figure~\ref{fig:ctmrgerror} (upper panels) shows the symmetry error for the  non-symmetric CTMRG as a function of $\chi$ and different values of $D$. For small bond dimension, the symmetry error vanishes for most values of $\chi$. An increasing number of non-vanishing errors is obtained at larger $D$ for specific values of $\chi$, but their magnitudes remain relatively small. 
 
Figure~\ref{fig:ctmrgerror} (lower panels) shows the difference in energy between the symmetric and non-symmetric CTMRG, $\Delta E = |E_{sym} - E |$. We find that a finite $S_{err}$ induces also a difference in  energy. Sizable deviations of the order $\Delta E \sim  10^{-4} - 10^{-3} $  (in units of the Heisenberg exchange coupling $J$)  can be found at large $D$ and small $\chi$. However, with increasing $\chi$, the observed deviations become substantially smaller~\cite{commentvariational}.

\begin{figure}[tb]
  \centering
  \includegraphics[width=\linewidth]{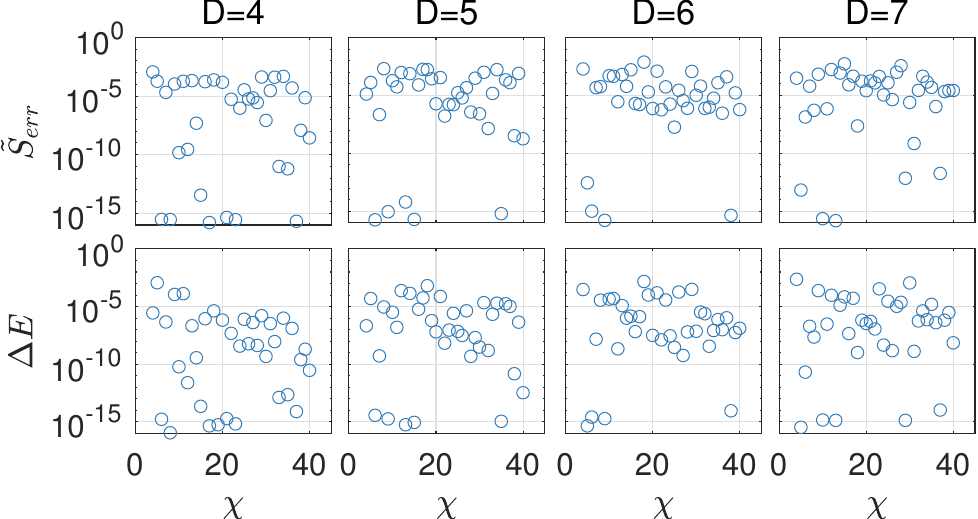}
  \caption{Upper panels: Symmetry error in HOTRG as a function of $\chi$ for different bond dimensions  $D$. Lower panels: Energy difference between the non-symmetric and $\mathbb{Z}_2$ symmetric implementation of HOTRG. }
  \label{fig:hotrgerror}
\end{figure}

In Fig.~\ref{fig:hotrgerror} we present similar results obtained with HOTRG. The maximal deviation $\Delta E$ is of similar order as in the CTMRG case, however, symmetry errors are found to occur more frequently.
One possible reason is that in CTMRG, a new bulk tensor $a$, which is perfectly symmetric, is absorbed into the environment at each iteration, which probably helps to preserve the symmetry in the environment up to a large extent. In HOTRG, however, only the initial tensor $a$ in the first iteration is perfectly symmetric, whereas at later iterations two coarse-grained tensors are combined, such that symmetry errors can accumulate.

In both cases, implementing the $\mathbb{Z}_2$ Hermitian symmetry enables to keep the tensors perfectly symmetric, i.e. the symmetry error is zero by construction.

\subsection{Computational speedup}
\label{sec:results}

\begin{figure}[tb]
  \centering
  \includegraphics[width=\linewidth]{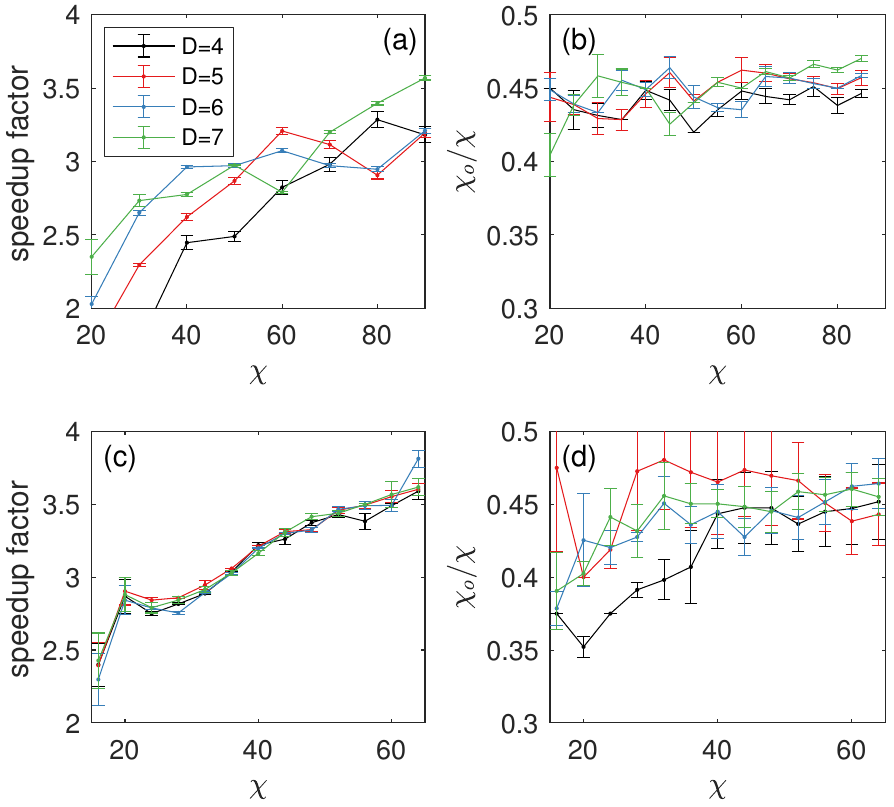}
  \caption{Speedup factor between non-symmetric and symmetric CTMRG (a) and HOTRG (c) as a function of $\chi$ for different bond dimensions $D$. The corresponding average ratio of the even and odd sectors with the corresponding standard deviation is shown in (b) and (d), respectively.  }
  \label{fig:speedup}
\end{figure}

In Fig.~\ref{fig:speedup}(a) we present results for the computational speedup for different values of $D$ and $\chi$, based on the average time of a CTMRG iteration. For small $D$ and $\chi$, the speedup factor is smaller than at large values. This is because the calculation with $\mathbb{Z}_2$ symmetric tensors comes with a certain overhead to permute and reshape the tensors. For larger tensors, this overhead becomes small compared to the time of the matrix-matrix multiplications or SVD, such that larger speedups can be achieved. This overhead could be further reduced by making use of a more efficient $\mathbb{Z}_2$ tensor class implementation. 

Another factor which influences the time are the block sizes of the even and odd sectors. In Fig.~\ref{fig:speedup}(b) we can see that the size of the odd sector is increasing with $\chi$, with a value of around $47\%$ for $D=7, \chi=100$ which is already close to optimal (i.e. 50\%). 
In Fig.~\ref{fig:speedup}(c),(d) we present similar results for HOTRG.  Also here we observe that the speedup factor approaches four with increasing~$\chi$. Here, the speedup factor is essentially independent of $D$, because $D$ is only relevant in the first iteration in HOTRG. \\

\section{Conclusions}
In this paper, we have shown how to incorporate  the Hermitian symmetry in 2D tensor network contractions. In the case of real-valued tensors, the double-layer tensors exhibit a $\mathbb{Z}_2$ symmetry which can be implemented using standard  symmetric tensor libraries. Exploiting this symmetry yields a speedup of up to a factor 4 in the large bond dimension limit, and the Hermitian symmetry of reduced density matrices is preserved by construction.

We also discussed the complex-valued case, where the tensors exhibit a block sparse structure for the real part and imaginary part separately, with the real (imaginary) numbers lying in the total even (odd) sector. Exploiting this structure reduces the number of operations by a factor 4, because tensor contractions involve multiplications between purely real numbers and/or purely imaginary numbers, instead of a multiplication of general complex numbers. 

Besides  CTMRG and HOTRG, our approach can be applied in various other context, e.g. in other 2D~\cite{levin07,orus2009-1,corboz14_tJ,evenbly15,zauner-stauber18,fishman18,homma24} or 3D contraction algorithms~\cite{vlaar21,vlaar23}, or to preserve the Hermiticity of density matrices in single-layer finite temperature algorithms~\cite{kshetrimayum19} or in open-system time evolutions~\cite{kshetrimayum17,mckeever21}.

\acknowledgments
This project has received funding from the European Research Council (ERC) under the European Union's Horizon 2020 research and innovation programme (grant agreement No. 101001604). J.H. acknowledges support from the Swiss National Science Foundation through a
Consolidator Grant (iTQC, TMCG-2 213805).

\bibliography{../bib/refs.bib,refs2.bib}

\begin{thebibliography}{75}
\bibitem [{su2()}]{su2comment}%
  \BibitemOpen
  \href@noop {} {}\bibinfo {note} {Similar issues arise in the SU(2) case,
  where transforming into the $\mathbb{Z}_2$ Hermitian basis would require to
  mix states $|S_{total}, S^z=k \rangle$ with $|S_{total}, S^z=-k \rangle$ in
  each spin sector $S_{total}$.}\BibitemShut {Stop}%
\bibitem [{com()}]{commentvariational}%
  \BibitemOpen
  \href@noop {} {}\bibinfo {note} {We note that $(E_{sym}-E)$ is not
  necessarily positive, because if the non-symmetric state is not Hermitian
  (i.e. unphysical), it is not strictly variational. Also note that in general
  the contraction error itself may not be strictly variational.}\BibitemShut
  {Stop}%
\end{thebibliography}

\end{document}